\documentclass[11pt,a4paper]{article}
\usepackage[authoryear]{natbib}
\usepackage{times}
\usepackage{graphicx}
\usepackage{epsfig}
\usepackage{amssymb,amstext}
\usepackage{amsmath}
\usepackage{setspace} 
\usepackage{authblk}
\usepackage{caption}
\usepackage{hyperref}
\usepackage{verbatim}
\usepackage{subcaption}
\usepackage{hyperref}
\usepackage[toc,page]{appendix}
\usepackage{bm}
\usepackage{float}
\usepackage[T1]{fontenc}

\usepackage[scale={0.7,0.85},centering,includeheadfoot]{geometry}

\def\E{\text{E}}
\def\mm#1{\ensuremath{\boldsymbol{#1}}} 
\title{Vole synchrony}
\date{}

\newpage
\begin{document}
\begin{titlepage}

\title{Seasonality, density dependence and spatial population synchrony}

\author{Pedro G. Nicolau\thanks{\textit{Address for correspondence}: Pedro Guilherme Nicolau, Department of Mathematics and Statistics, Faculty of Science, UiT The Arctic University of Norway, 9037 Troms{\o}, Norway. \E-mail: pedro.nicolau@uit.no} $^1$, Sigrunn H. S{\o}rbye$^1$, Rolf A. Ims$^2$ and Nigel G. Yoccoz$^2$}
\affil{$^1$Department of Mathematics and Statistics, Faculty of Science, UiT The Arctic University of Norway}
\affil{$^2$Department of Arctic and Marine Biology, Faculty of Biosciences, Fisheries and Economics, UiT The Arctic University of Norway}

\date{}
\end{titlepage}
\maketitle
\clearpage

\section*{Abstract}
Spatial population synchrony has been the focus of theoretical and empirical studies for decades, in the hopes of understanding mechanisms and interactions driving ecological dynamics. In many systems, it is well-known that seasonality plays a critical role in the density-dependence structure of the populations, yet this has hardly received any attention in synchrony studies. Here, we propose a protocol that allows to elucidate deterministic and stochastic sources of spatial synchrony, while accounting for geographic- and season-specific density dependence. We apply our protocol to seasonally-sampled time series of sub-arctic gray-sided voles, known for marked spatial synchrony. Dissociating seasonal density-dependence contributions to the total observed synchrony reveals differential strength and shape of synchrony patterns by season. Mild winter weather reveals to be an important driver of vole spatial synchrony, with lagged effects in the fall. This has direct implications to the future population dynamics of such species when facing climate change.

\section{Introduction}

Spatial synchrony -- referring to the extent local populations display simultaneous changes across space -- is a universal characteristic of geographically distributed populations. The strength and scale of population synchrony, which varies tremendously between species and ecosystems, has been the subject of a large number of theoretical and empirical studies (reviewed by \cite{Liebhold2004, Hansen2020}). These studies are motivated by their potential to provide unique insights into the mechanisms that drive ecological dynamics across a range of spatial scales \citep{Bjornstad1999a, Koenig1999, Walter2017}. The study of spatial population synchrony is one of the fields within ecology that is, both conceptually and methodologically, most tightly linked to other sciences that also deal with spatio-temporal dynamics \citep{Nareddy2020, PerezGarca2021}.

P.A.P. \citet{Moran1953} developed the first formal theory of spatial population synchrony.  Moran’s theorem postulates that populations subjected to the same regulatory biotic mechanisms (i.e. log-linear density dependence), and influenced by the same (or perfectly correlated) abiotic environmental variation (e.g. stochastic weather), will display a synchrony that mirrors the synchrony of the environmental variation \citep{Moran1953,Hudson1999,Hansen2020}. While this theorem has become a cornerstone of the study of population synchrony, Moran himself expressed the need for relaxing some of its restrictive assumptions in order to be more applicable to empirical case studies. Subsequently, many studies have contributed to a “generalization of the Moran effect” (sensu \cite{Hansen2020}) by, for instance, allowing for non-linear density dependence \citep{Blasius1999,Engen2005}, spatially heterogeneous \citep{Royama2005,Hugueny2006} and temporally autocorrelated environmental variation \citep{Massie2015}, and inclusion of other synchronizing mechanisms (e.g. dispersal \cite{Ripa2000} and trophic interactions \cite{Jarillo2020}). Analytical approaches to elucidate the effect of climatic variation on population synchrony have become particularly timely in the current era of anthropogenic climate change \citep{Sheppard2015, Koenig2016}.

Accounting for seasonality was a fundamental aspect highlighted by \citet{Moran1953} when assessing the effect of meteorological conditions on population synchrony. This became clear to him when analyzing population time series of lynx from boreal Canada, which is a region with strikingly different climate in summer and winter. Moran realized that season-specific biotic mechanisms were important, because different demographic parameters are involved in the two seasons (e.g. reproduction only in summer). However, because the lynx population time series were based on only one census per year, Moran was not able to analytically account for season-specific population processes (i.e. density dependence). More modern studies of seasonally-sampled boreal and arctic rodent populations have shown that marked season-specific density dependence is indeed present and a crucially important determinant of local population dynamics \citep{Hansen1999, Stenseth2003, Fauteux2021}. Although seasonality is such a critical aspect of most ecological systems \citep{White2020}, and changing seasonality is one of the most profound consequences of global warming in the northern hemisphere \citep{Xu2013}, we are not aware of any study of population synchrony that has explicitly incorporated seasonality. On the contrary, it has even been argued that one should use yearly averages to remove the influence of seasonality when estimating synchrony (e.g. \cite{Dallas2020}).

The purpose of the present study is to devise a general stepwise analytical protocol, that accounts for season-dependent and geographic context-dependent population processes, to identify which aspects of climatic variation and change are most influential to spatial population synchrony (Fig.~\ref{fig:protocol}). We illustrate the applicability and potential of the protocol through a case study of the gray-sided vole (\textit{Myodes rufocanus}). This boreal-arctic rodent species is renowned for its important role in ecosystem functioning \citep{Boonstra2016} and multi-annual population cycles \citep{Hansen1999, Turchin2000}, with suspected impacts of climate change on these cycles \citep{Ims2008, Cornulier2013}.

\section{Results}
Local gray-sided vole abundances were estimated every spring and fall over 21 years based on capture-recapture sampling in northern Norway (Fig.~\ref{fig:datacycles} a, b). Nineteen sampling locations (i.e. live-trapping grids) were spaced along a 170 km transect in boreal mountain birch forest and encompassed three predefined geographic regions ($R_1$: coast, $R_2$: fiord and $R_3$; inland; Fig.~\ref{fig:datacycles}a) which were expected to influence the density dependent structure of vole population dynamics.  

The 21-year population time series encompass five multi-annual cycles, exhibiting profound overall synchrony across the extent of the study area (Fig.~\ref{fig:datacycles}b). However, despite visible spatial synchrony, and relative temporal stationarity, there is also some variation in timing and amplitude of the cyclic peaks among the localities. This regards especially the spring series, which have lower and more variable abundance estimates than the fall series (Fig.~\ref{fig:datacycles}b). 

Previous studies have demonstrated that local boreal and Arctic vole populations are adversely affected by winter weather phenomena, such as thaw-freeze cycles \citep{Aars2002,Kausrud2008} and rain-on-snow events \citep{Fauteux2021}. Hence, we derived local time series of the number of days the temperature crossed zero degrees (Celsius), and the total amount of rainfall (mm) during winter (Fig. \ref{fig:datacycles}d). The two weather variables exhibit spatial synchrony, with a tendency for milder (more zero crosses) and wetter (more rainfall) climate towards the coastal area.

\subsection{Density-dependence structure}
Following \citet{Stenseth2003}, we fitted second-order log-linear autoregressive models to the population time series, according to the density-dependent (DD) models described in Fig.~\ref{fig:protocol} (models II-IV). As we use a Bayesian framework to conduct the data analysis, we selected Bayesian $R^2$ \citep{Gelman2019} as a measure of explained variance (i.e. the fit) of the different linear autoregressive models. In general, the models explained more of the abundance variance in the fall than the spring (Fig.~\ref{fig:r2}). The inclusion of geographic region-specific DD parameters (when comparing model II and III; Fig.~\ref{fig:r2}) did not improve the model fit much, suggesting there are small differences in the DD structure between the three geographic regions. However, a large improvement of the model fit was achieved when including including season-specific DD parameters (model IV; Fig.~\ref{fig:r2}), especially concerning the fall abundances. This implies that season-specific biotic interactions are strongly influential components of the overall population dynamics.    

\subsection{Spatial population synchrony}
The spatial correlograms, based on the four population metrics (I-IV) outlined in Fig. 1, clearly show that much of the overall spatial population synchrony (Fig. \ref{fig:scale}, I) is due to a common DD structure across the study area (Fig. \ref{fig:scale}, II-IV). Moreover, when accounting for season-specific DD (model IV), the synchrony in the residuals drops substantially in comparison to that from models II and III, which only account for annual DD (Fig. \ref{fig:scale}, IV). The reduction in spatial synchrony due to seasonal DD is particularly sharp for the fall abundances, for which the synchrony between the most distant populations approaches zero. Accounting for the slight differences in density dependence among the three geographic regions provides almost no contribution to the synchrony pattern (i.e., comparison between II and III in Fig. \ref{fig:scale}).

\subsubsection{Weather synchrony vs. population synchrony}
The synchrony of both of the weather variables declined steeply as a function of distance between the sampling stations. However, there was more scattering in the cross-correlations in rainfall when compared to the correlations in the zero crosses (Fig. \ref{fig:weathersync}a). The synchrony of number of zero crosses was positively and significantly associated to population synchrony corrected for DD structure (model IV) both in fall and spring, while the synchrony in winter rainfall was only related to the population synchrony in the fall (Fig. \ref{fig:weathersync}b,c). 

\section{Discussion}
We have here proposed and exemplified an analytical protocol, that based on time series data, allows for elucidating deterministic and stochastic sources of spatial population synchrony. Potential deterministic sources include density dependence, climatic seasonality and geographic ecological context, while influential stochastic sources are likely weather variables. Spatial covariance in stochastic weather events amounts to the Moran effect provided that the deterministic components of local population dynamics are linear and identical. Nonetheless, under most circumstances, correlated weather events are expected to exert synchronizing effects when the local density-dependent structure is non-linear and spatially heterogeneous (i.e., the generalized Moran effect; cf. \cite{Engen2005,Royama2005,Hansen2020}).

\citet{Moran1953} showed that a key step to make “meteorological phenomena show up more clearly” in statistical analyses of population synchrony is to remove the density-dependent structure from the population time series before making further statistical inferences (e.g., by analyzing the residuals of an autoregressive model). Many studies have used Moran’s approach to remove serial autocorrelation in order to fulfill the independence requirement for significance tests of synchrony \citep{Buonaccorsi2001, Liebhold2004}. However, there appears to be a lack of studies that have followed Moran’s suggestion to formally analyze whether the scale of synchrony in the population residuals is dependent on synchrony in the weather (but see \cite{grotan2005climate}); i.e., as achieved by step IV in our analytical protocol. Accordingly, \citet{Hansen2020} conclude that there has been an “analytical deficiency” in empirical Moran-effect studies in terms of making formal inferences about how population synchrony is environmentally forced. We show in the present study that by focusing on residuals which by definition depend on an adequate model structure, we draw more accurate inferences regarding the strength and scale of synchrony.

By applying our analytical protocol to bi-annually sampled time series of gray-sided vole populations we demonstrate winter weather contributions to spatial synchrony. We found that both the amount of rainfall and the frequency of mild-spells in winter contribute to spatial synchrony. These two weather variables have previously been found to affect local population dynamics of boreal and arctic vole species by enhancing winter declines \citep{Aars2002,Fauteux2021}. However, the present study is the first to analytically link large-scale spatial synchrony -- a phenomenon that appears to be ubiquitous in boreal and arctic cyclic small rodent populations (cf. \cite{Stenseth1993,Krebs2013}) -- to any form of stochastic environmental forcing; i.e. Moran effects.

An interesting result arising from our analysis is the time-lagged effect of the winter weather on synchrony of fall abundances. \citet{Moran1953} found similar time-lagged weather effects on an annual time-scale for Canada lynx and speculated about which biological mechanisms could be involved. In voles, environmental conditions in the non-breeding seasons may have lasting effects, for instance, by delaying the onset of reproduction and thereby reducing population growth over the summer \citep{Ergon2001}.  The combination of direct and lagged effects of winter weather amounts to an enhanced Moran effect. As increased frequencies of rain-on-snow events and thaw-freeze cycles are very likely outcomes of climate warming in boreal and Arctic ecosystems \citep{AMAP2017}, we predict that the strength and scale of spatial synchrony of rodent populations will change in these ecosystems.    

Climatic seasonality is an externally forced oscillator that acts on the dynamics of most natural systems \citep{Fretwell1972}. Yet both empirical and theoretical studies of ecological dynamics mostly ignore this fact \citep{White2020}. While seasonality has been shown to be a very important component of spatio-temporal disease dynamics \citep{Earn1998,Grenfell2001,Moustakas2018}, we are not aware of empirical studies that have explicitly investigated how such seasonal forcing acts on the strength and scale of synchrony in animal population dynamics. Our analytical protocol provides means for filling this knowledge gap. Specifically, the role of seasonality becomes evident by comparing the correlograms of residuals from models with and without seasonal density dependence (i.e. compare correlograms III and IV in Fig.~\ref{fig:scale}). In the case of sub-arctic gray-sided voles, seasonality is evidently an important determinant of the region-scale spatial synchrony. This regards especially the fall abundances, for which both the overall synchrony becomes reduced and the distance effect is enhanced when seasonal density dependence is accounted for. In this case, it appears that the exact nature of such season-specific effects is contingent on the relative magnitude of the spring and fall noise term of the bi-variate autoregressive model (see Appendix \ref{app:Simulations}).

The role of seasonality may be a particularly forceful determinant of spatio-temporal population dynamics in species with multivoltine life cycles, like voles. For instance, the length of winter seasons has been found to exert a strong effect on the local vole population dynamics by acting through density dependent structure \citep{Batzli1999,Stenseth1999,Stenseth2003,Bierman2006} and likely also through season-specific noise terms (\cite{Vasseur2007}, Appendix \ref{app:Simulations}). Hence, it may not be surprising that seasonality also exerts an effect on regional population dynamics (e.g. large-scale spatial synchrony) as here shown for sub-arctic gray-sided voles. However, as demographic processes are typically season-specific also in univoltine species \citep{Boyce1999} -- including how they are affected by density-dependent and independent factors --, we believe that our analytical protocol (Fig. \ref{fig:protocol}) will help advance empirical studies of spatial population synchrony for a wide range of species.

\section{Methods}
The methods section follows the structure outlined in Fig.~\ref{fig:protocol}, with the five steps required to investigate weather effects on the spatial population synchrony of a gray-sided vole population, after accounting for the geographical- and seasonal-DD structure of the population. 
\subsection{Sampling and Time Series (Steps 1 and 2)}
\subsubsection{Data and Study Area}
We use data from a long-term running monitoring program of the rodent community in the region of Porsanger, northern Norway, between 2000-2020. The data collection consisted of a capture-mark-recapture methodology with two trapping days at 19 individual stations, scattered along a linear transect of approximately 170 km of road. Trapping sessions were conducted twice per year, once in late spring after snow melt, and once at the end of the summer, at the end of the vole reproductive season  (see \citet{Ehrich2009} for precise trapping specifications).
The Porsanger region contains different landscapes and is subject to a strong climatic contrast (in both temperature and precipitation). The different stations can be sorted into $m=3$ regions according to their landscape affinities: coastal region ($R_1$), fjord region ($R_2$) and inland region ($R_3$). Stations 1--5 were included in $R_1$ ($n_1=5$), stations 6--12 were included in $R_2$ ($n_2=7$) and stations 13--19 were included in $R_3$ ($n_3=7$). Fig.~\ref{fig:datacycles} summarizes spatial features of the study area and data. 

\subsubsection{Abundance estimation from mark-capture-recapture data}
To reduce a potential bias when estimating synchrony \citep{SantinJanin2014}, we incorporated the sampling error from capture heterogeneity in our estimates of seasonal abundances \citep{Nicolau2020}. Specifically, we fitted a multinomial regression model to the capture history data to estimate the probability of obtaining a given capture history as a function of individual features registered during the live trapping. These features included the \textit{weight} and \textit{sex} of the individuals. We also added a random effect for station in the predictor of the regression model. Individual capture probabilities were subsequently estimated by assuming a temporal effect on the capture process (model $M_{th}$, \citet{Otis:1978}). Finally, the individual probabilities were used to estimate seasonal abundances using an empirical Horvitz-Thompson estimator, which is a function of the estimated individual capture probabilities.
Denote the resulting estimated log abundances by  $\{X_{s,t}\}$ and $\{Y_{s,t}\}$, for spring and fall, respectively, at spatial locations $s=1,\ldots , n_s$ and year $t=1,\ldots , n_t$. For the case study, $n_s = 19$ and $n_t = 21$.

\subsubsection{Weather variables}
To explore the effect of the weather on the spatial synchrony, we should ideally look into the winter snow conditions (i.e, snow depth and ice formation) as they can be considered the most relevant climatic variables affecting rodent population cycles \citep{Hansson1985,Hansen2013,Stien2012,Fauteux2021}, being an interaction of different weather variables, including temperature and precipitation. As this information was not directly available, we resorted to proxy variables of snow conditions, using the temperature and precipitation estimates from the Norwegian Meteorological Institute between 2000-2020. These measurements correspond to model estimates (not measured at station level) and are prone to large uncertainties, particularly the precipitation \citep{Lussana2019}. For our proxies of snow conditions, we derived two variables: winter \textit{zero crosses}, as the total number of times the mean daily temperature crossed 0 °C during winter (21 Dec -- 20 Mar); and \textit{winter rainfall}, as the precipitation sum in days where the mean temperature surpassed 0 °C, during winter.

\subsection{Statistical Framework}
We describe the statistical framework to decompose density-dependence contributions into the spatial synchrony of populations, and isolate the weather effects on population synchrony, described in steps 3--5 in Fig.~\ref{fig:protocol}.

\subsubsection{Density-dependence structure (Step 3)}\label{sec:dd}
The general protocol (Fig. 1, step 3) specifies three different models for the DD structure of the estimated time series. Here, we assume that 
the general function $f(.)$ is linear, 
describing the log-DD structure in terms of direct and delayed effects up to lag $p$.  Specifically, the three models either include or exclude regional- and seasonal-dependent effects as specified below. 

\begin{figure}[!h]
    \centering
    \includegraphics[width=1\textwidth,keepaspectratio]{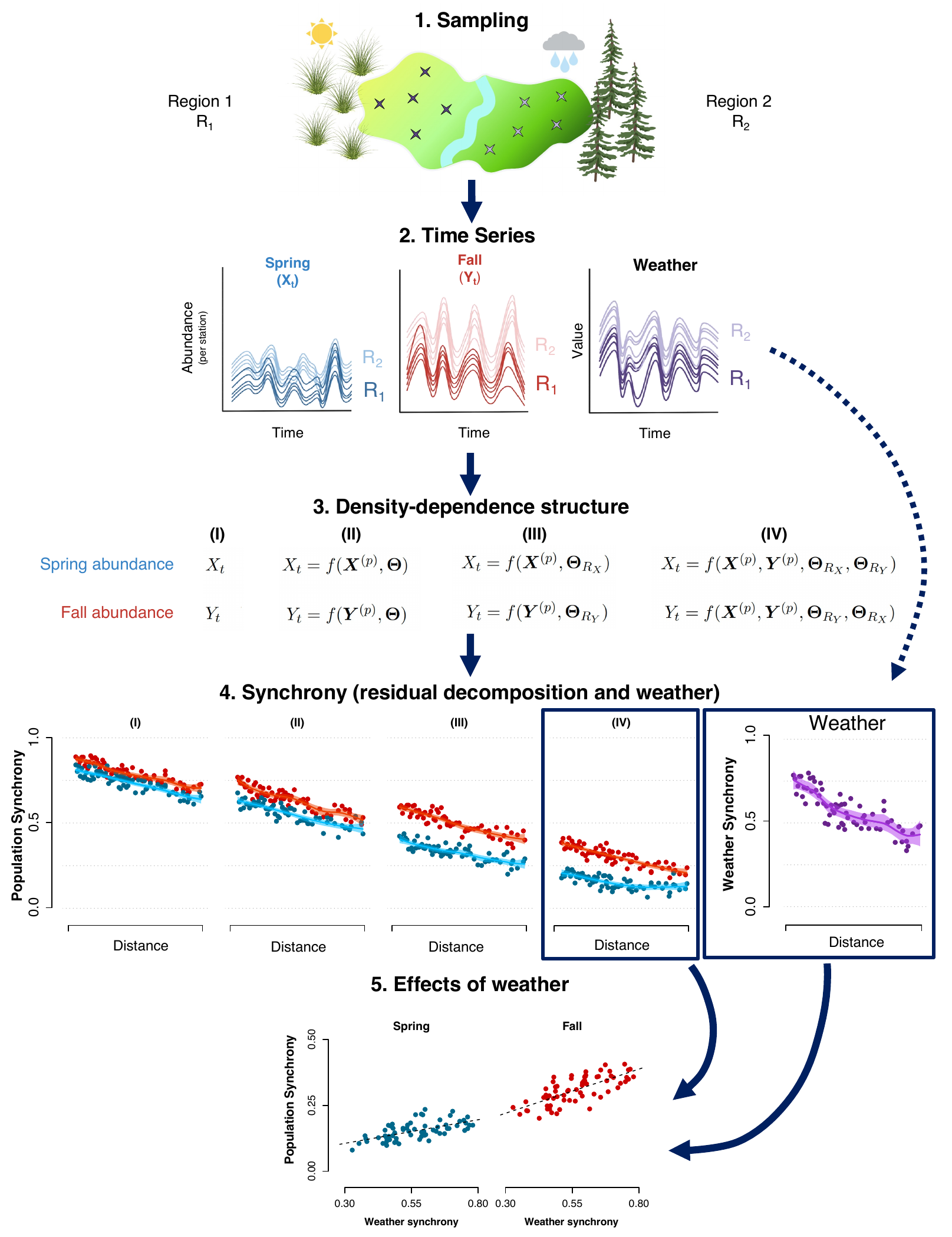}
    \caption{Methodological Protocol (cont.)}
    \label{fig:protocol}
\end{figure}

\begin{figure}[H]
    \centering
    \includegraphics[width=1\textwidth,keepaspectratio]{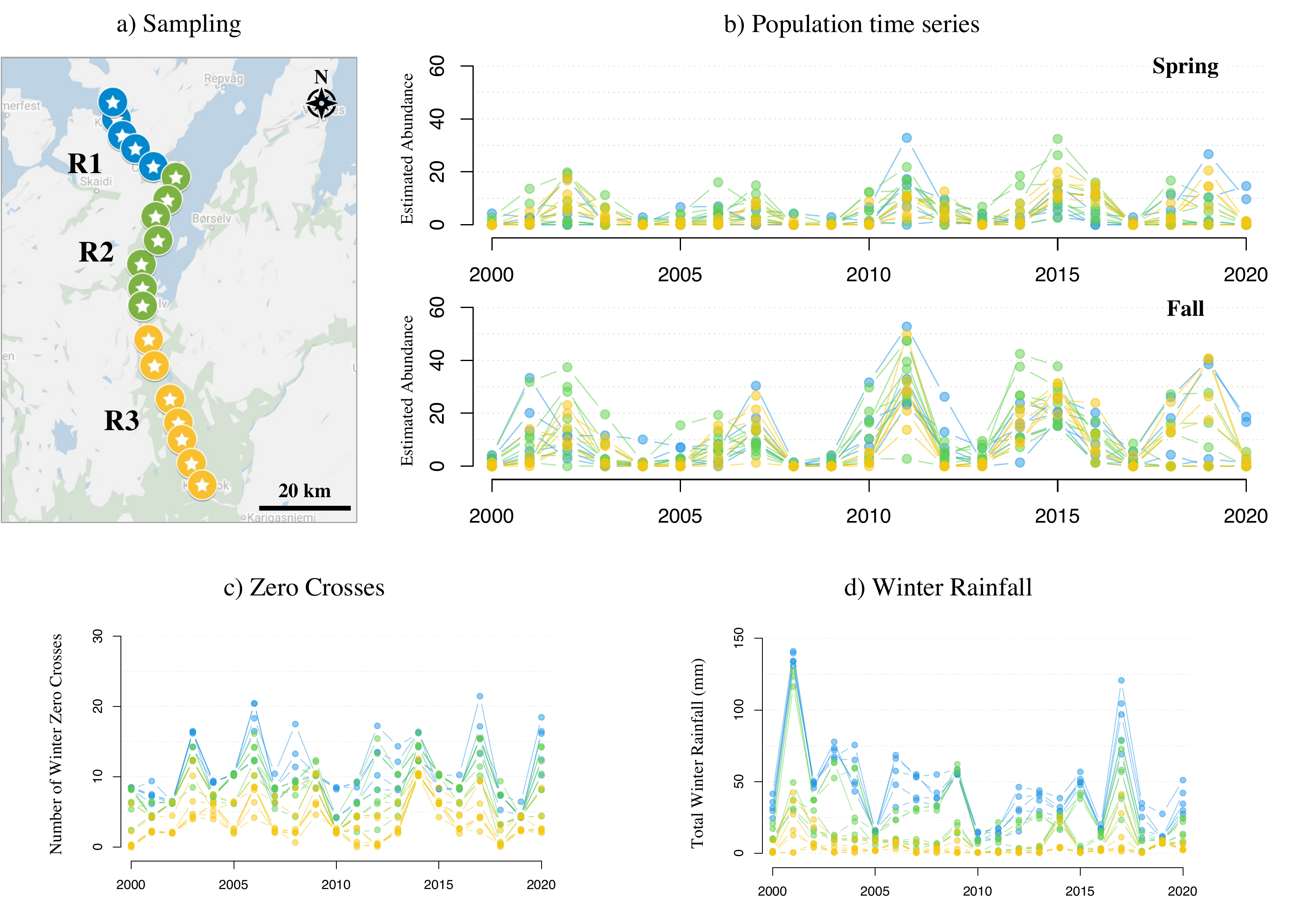}
    \caption{Sampling design and time series. a) Map with the 19 sampling stations (“dot/stars”) along the 170 km transect encompassing three geographical contexts ($R_1$: coast, $R_2$: fjord; $R_3$: inland). Green shade landcover denotes mountain birch forest, white denotes tundra and blue denotes water surfaces. b) Time series of abundance estimates for the 19 local grey-sided vole population in spring (lower) and fall (upper). Time series of the two focal winter weather variables are presented in c) number of zero crossings and d) total winter rainfall.  Colors of curves in b)--d) correspond to the three geographic context in a). }
    \label{fig:datacycles}
\end{figure}

\begin{figure}[H]
    \captionsetup[subfigure]{labelformat=empty,aboveskip=0pt,justification=centering}
  \centering
    \includegraphics[width=.65\linewidth]{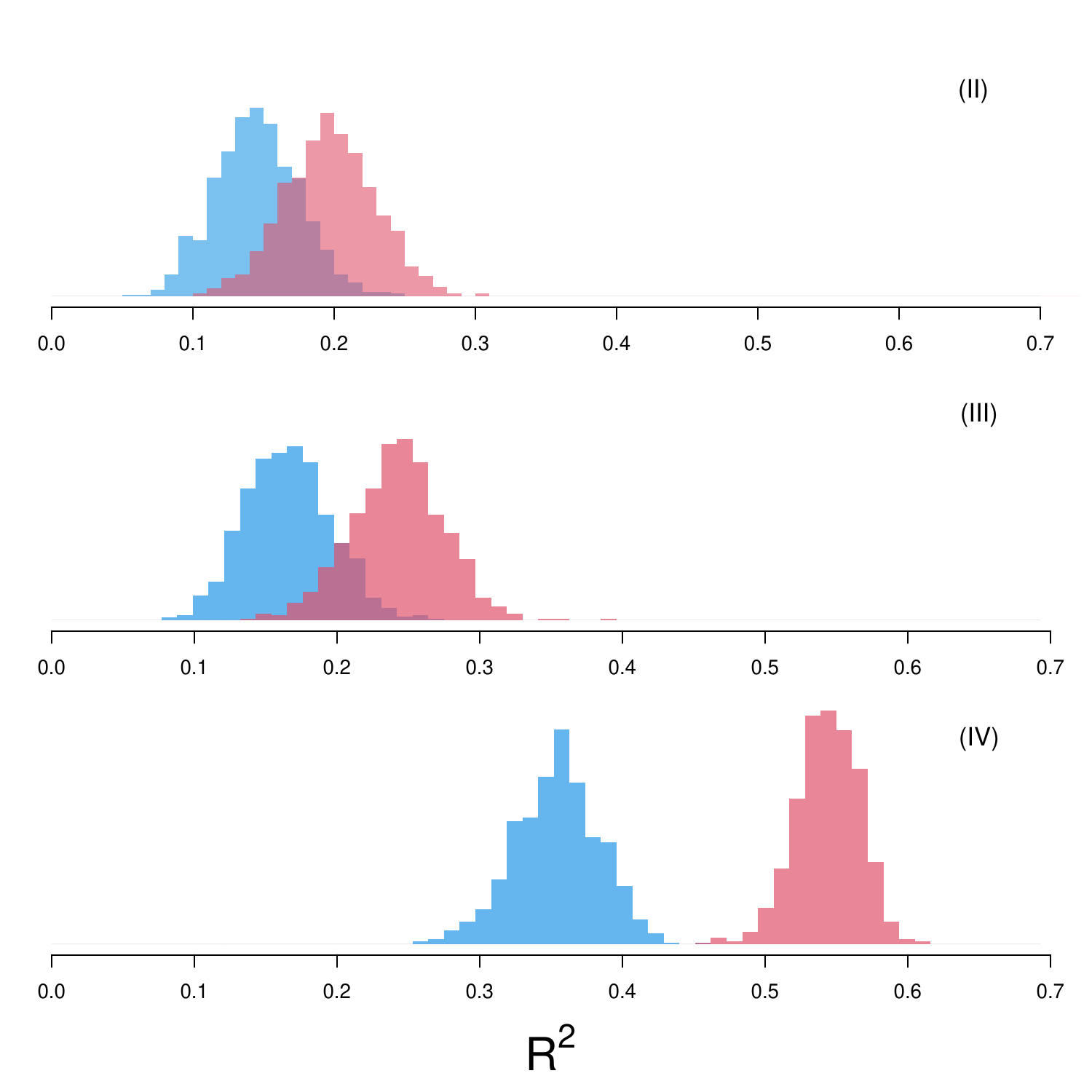}
    \caption{Distributions of Bayesian $R^2$ values (blue for spring and red for fall abundances) for the three sets of second-order, log-linear models corresponding to the general DD models II –- IV outlined Step 3 Fig. 1. As we are working in a Bayesian framework the computation of the $R^2$ results in a distribution itself (see \cite{Gelman2019} for details). }
  \label{fig:r2}
 \end{figure}

\begin{figure}[H]
    \centering
    \includegraphics[width=.7\textwidth,keepaspectratio]{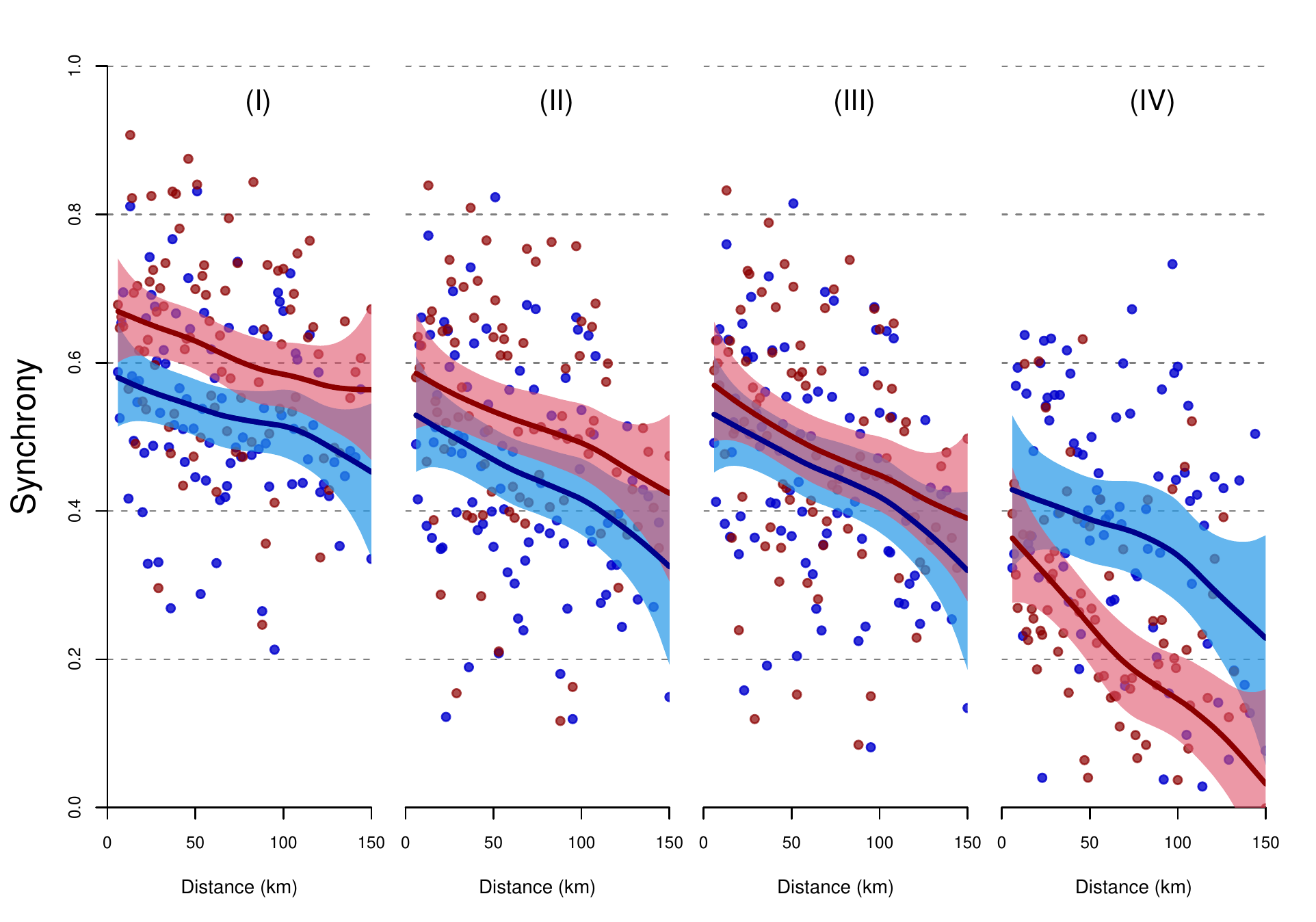}
    \caption{Correlograms depicting how the synchrony (i.e. the pairwise spatial correlations) of four different population metrics (see Fig. \ref{fig:protocol}, Step 3) varies with geographical distance between local gray-sided vole populations. 95\% credible intervals for spring are plotted in blue, and for fall are plotted in red, while the solid lines correspond to the median. Single dots correspond to the individual pairwise correlations used in the correlogram.}
    \label{fig:scale}
\end{figure}

\begin{figure}[H]
    \centering
    \includegraphics[width=1\textwidth,keepaspectratio]{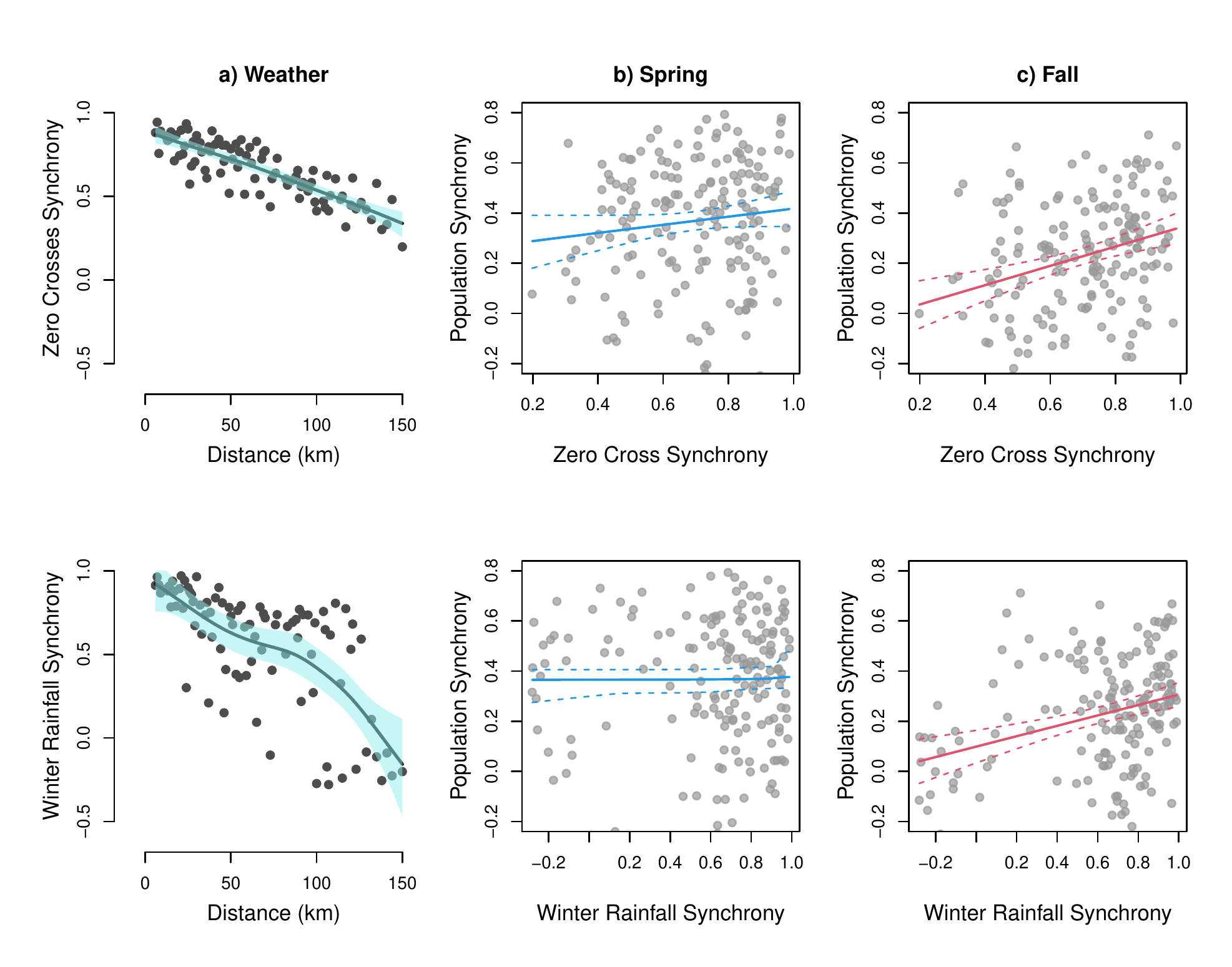}
    \caption{Weather synchrony versus population synchrony. Panels a) Correlograms of the two focal  weather variables with associated 95\% credible intervals, winter \textit{zero crosses}  (ZC; top) and winter rainfall (WR; bottom). Panels b) and c) correspond to linear regression lines, with associated 95\% credible intervals, of population synchrony as a function of weather synchrony for spring (b) and fall abundances (c). Slope estimates for spring are $\beta_{ZC} = 0.16$ (CI : -0.04,0.36) $\beta_{WR} = 0.00$ (CI : -0.11,0.10). Slope estimates for fall are $\beta_{ZC} = 0.38$ (CI : 0.20,0.56) and $\beta_{WR} = 0.21$ (CI : 0.12,0.30). CI denotes 95\% credible intervals for each regression coefficient $\beta$.  
    }
    \label{fig:weathersync}
\end{figure}

In general, the spatial locations $s$ are assumed to be within a closed geographical region $R$, which can be partitioned into $m$ mutually exclusive subregions, $R = R_1\cup \ldots \cup R_m$. For the gray-sided vole case study, this corresponds to $m=3$ regions. The most general model includes both regional-specific and seasonal-specific terms (Fig.~\ref{fig:protocol}, model (IV)), and the assumed log-linear dependency structure up to order $p$ can be expressed by 
\begin{eqnarray}
X_{s,t}  &=&  \beta_{r1}Y_{s,t-1} + \beta_{r2} X_{s,t-1} +\ldots + \beta_{r,2p-1} Y_{s,t-p}  +\beta_{r,2p}X_{s,t-p}+ \epsilon_{s,t} \label{eq:winter}\\
 Y_{s,t}  &=&  \gamma_{r1}X_{s,t}  + \gamma_{r2} Y_{s,t-1}+ \ldots + \gamma_{r,2p-1} X_{s,t-1} + \gamma_{r,2p}Y_{s,t-p} + \omega_{s,t} \label{eq:summer}
\end{eqnarray}
where $t=p+1,\ldots , n_t$ and $s\in R_r$.  The terms $\epsilon_{st}$ and $\omega_{st}$ denote individual random environmental noise at each spatial location $s$ for each time point $t$, while the sets of regional- and seasonal-specific coefficients can be summarized as $\Theta_{R_x} = \{\beta_{r1},\ldots ,\beta_{r,2p}\}$ and  $\Theta_{R_y}= \{\gamma_{r1},\ldots ,\gamma_{r,2p}\}$.

Simplifications of the given model will yield more simplistic measures of the DD structure. According to the general protocol in Fig.~\ref{fig:protocol}, Model (I) corresponds to assuming no DD structure, in which all of the given coefficients are equal to 0. This corresponds to simply using the estimated raw log-abundance series, $\{X_{s,t}\}$ and $\{Y_{s,t}\}$, in further analysis. 

Following \citep{Stenseth2003}, we included delayed effects up to order $p=2$ for the case study. Model (II) refers to a second-order annual autoregressive processes including coefficients $\Theta = \{\beta_2, \beta_4, \gamma_2, \gamma_4\}$  which are neither regional-specific ($m=1$; disregarding spatial heterogeneity), nor seasonal-specific ($\beta_1=\beta_3=\gamma_1=\gamma_3=0$; assuming yearly dynamics). Such AR(2) models are often used in literature (e.g. \cite{Turkia2020, Dallas2020}).
Model (III) is characterized by incorporating regional-specific effects $\{\beta_{r2}, \beta_{r4}, \gamma_{r2}, \gamma_{r4}\}_{r=1}^m$. This corresponds to AR(2) models which allow for spatial differences in the DD structure which can account for some of the observed synchrony \citep{Hugueny2006}. Finally, by including the seasonal-specific effects  $\{\beta_{r1}, \beta_{r3}, \gamma_{r1}, \gamma_{r3}\}_{r=1}^m$, we get the bivariate model (IV)  which is very similar to a second-order vector autoregressive model (VAR). The difference to a VAR-model, however, is that the time series $\{X_{s,t}\}$ and $\{Y_{s,t}\}$ are observed at two different time points in year $t$, and the fall log-abundances are modeled in terms of the spring observations within the same year. Seasonal-specific DD has been recognized as fundamental to model small rodent population dynamics \citep{Hansen1999}, but to our knowledge seasonal DD contributions to spatial synchrony have not been assessed.  

\subsubsection{Measuring the scale and shape of spatial population synchrony (Step 4)}\label{sec:synchrony}
To assess the scale and shape of the spatial synchrony, we can consider the spatial correlations of the environmental noise terms in Models (I--IV) as a function of geographical distance. The following analysis is repeated using the four different models for DD structure, specified in section~\ref{sec:dd}. A major goal is then to understand how the inclusion of regional- and seasonal-specific terms influences the synchrony estimates, i.e., which part of the synchrony is explained by the different DD components.

Define the residual vectors $\mm{\epsilon}_s' = (\epsilon_{s,1},\ldots , \epsilon_{s,n_t})$ and  $\mm{\omega}_s' = (\omega_{s,1},\ldots , \omega_{s,n_t})$ for all spatial locations $s=1,\ldots ,n_s$ and $t=p+1,\ldots , n_t$. The contributions to the spatial synchrony are then characterized by the pairwise correlations between vectors within each of the sets $\{\mm{\epsilon}_s'\}_{s=1}^{n_s}$ and $\{\mm{\omega}_s'\}_{s=1}^{n_s}$. If the associations between these residual series are expected to be linear, the degree of synchrony is typically measured using Pearson's correlation coefficient \citep{Bjornstad1999a, Liebhold2004}. To model the correlations in terms of geographical distance, let $\delta_{i,j}$ denote the Euclidean distance between two stations $i$ and $j$.  In accordance with calculating the spatial correlogram \citep{Bjornstad1999a, Bjornstad2001,Liebhold2004}, we discretize the $n_s(n_s-1)/2$ unique distances between stations into distance classes $d_k$, $k=1,\ldots , K$, where $K$ is the total number of classes. Specifically, a distance class $d_k$ is defined by $L_k<\delta_{i,j}<U_k$, where $L_k$ and $U_k$ represent the lower and upper bound of the distances within that class, respectively. The corresponding averages of the pairwise correlations $\{\rho_{i,j}\}$ for distance class $d_k$ are then given by 
\begin{equation}
    \rho_k(d_k) =
    \frac{2\sum_{i=1}^{n_k}\sum_{j=i+1}^{n_k}\rho_{i,j}}{n_k(n_k-1)},\quad L_k<\delta_{i,j}\leq{U_k},
    \label{eq:rho-average}
\end{equation}
\noindent 
where $n_k$ is the total number of distances/correlations within distance class $d_k$. The given formulation is analogous to the calculation of Koenig's modified correlogram \citep{Koenig1999,Bjornstad1999a}, as the correlations are not centered (zero synchrony is taken as the reference line of the correlogram). For the given case study, we assumed that the distance-class width is $U_k-L_k=1$ for all classes, which corresponds to rounding off the geographical distances to the nearest integer. We used this method to calculate the averaged correlations in \eqref{eq:rho-average} as a pre-processing step to reduce random noise in the estimated correlations.   

As an alternative to using the non-parametric covariance function  \citep{Bjornstad2001} or other non-parametric estimates of the correlation function \citep{Liebhold2004}, we chose to model the correlations in terms of the distances using the regression model 
\begin{equation}
\rho_{k}(d_k) = f(d_{k}) + \nu_{k},\quad k=1,\ldots , K.
\label{eq:smooth-f}
\end{equation}
Here, $f$ denotes a smooth underlying function while $\{\nu_{k}\}$ represents zero-mean, independent Gaussian error terms with constant variance. This model is fitted using a Bayesian framework where the function $f$ is assigned a second-order intrinsic Gaussian Markov random field prior (\cite{Rue:2005}, page 110). The model is scaled according to \citet{sorbye:14} and the precision parameter of the model is assigned a penalized complexity prior with parameters $U=0.5$ and $\alpha=0.01$ \citep{simpson:17}. Using the methodology of integrated nested Laplace approximation \citep{Rue:2009}, both the posterior mean and credible intervals for $f$ are calculated efficiently, without the need of resampling techniques, like Monte Carlo simulation or bootstrapping.

\subsubsection{Effects of Weather (Step 5)}
Finally, we can use the measures of synchrony accounting for the effects of geographic- and seasonal-dependent DD to investigate potential weather (or other relevant environmental variables) drivers. For this, we can model the set of correlations $\{\rho_k(d_k)\}_{k=1}^K$ from model (IV) as a function of the corresponding spatial correlations of different weather covariates, defined by $\{\rho^{(c)}_k(d_k)\}_{k=1}^K$. The availability of such covariates are typically case-specific but should be  measured or estimated to represent the same spatial locations and time points used for the log-abundance estimates.  For the given case study, the relationship between the weather variables (\textit{zero crosses} and \textit{winter rainfall}) appeared to be linear, and was thus modeled using simple linear regression models.

\newpage
\section*{Figures}

\newpage

\noindent Figure 1: (cont.) The five main steps of a general methodological protocol to single out the impacts of climatic variation (weather) on spatial population synchrony, by accounting for seasonal and geographical contexts in ecological population processes (density dependence). 
Step 1: Seasonal (spring and fall) sampling of both local populations and a focal weather variable at different locations (crosses). The geographic sampling frame encompasses two regions ($R_1$ and $R_2$) representing different geographic ecological contexts (e.g. habitats or ecological communities). 
Step 2: Season- and region-specific time series of local population abundance estimates resulting from the sampling process, together with time series of the focal weather variable. The estimation of abundance ideally involves separating the observation process and the population process, accounting for detectability.
Step 3: Four alternative models to further analyze spatial population synchrony. (I) corresponds to seasonal abundance estimates ($X_t$ and $Y_t$). (II-IV) correspond to the sets of $X_t - f(.)$ and $Y_t - f(.)$ from the respective general models for density-dependence, modeling state of the population at time $t$ as a function of previous $p$ states. Model (II) includes only one set $\Theta$ of density-dependence parameters with annual time lags (i.e. ignoring seasonal and regional components). Model (III) includes region-specific parameters $\Theta_{R}$, again with annual time lags (i.e. ignoring seasonal components). Model (IV) is a bi-variate model \citep{Stenseth2003} that includes both geographic- and season-specific parameters $\Theta_{R_X}$ and $\Theta_{R_Y}$. 
Step 4: Season-specific synchrony patterns (i.e. scale and shape) of the population (derived from Step 3) and weather metrics (derived from Step 2) as function of distance. The dots are the pairwise cross-correlations of the population metrics and the weather variables, while the lines are estimated correlograms with associated uncertainty intervals (e.g., \citet{Bjornstad1999a}).
Step 5: Estimated effects of weather synchrony on population synchrony. Season-specific (Fall and Spring) population synchrony with corrected for seasonal-density dependence and geographic context effects (i.e. residuals from model (IV)) are regressed against the spatial synchrony in the focal weather variable. Illustrations created with Biorender.com.

\newpage
\clearpage

\newpage
\pagebreak
\bibliographystyle{apalike}
\bibliography{synchrony}





\newpage
\appendix
\section{Winter Weather Patterns}

\begin{figure}[!ht]
    \centering
    \includegraphics[width=\textwidth,keepaspectratio]{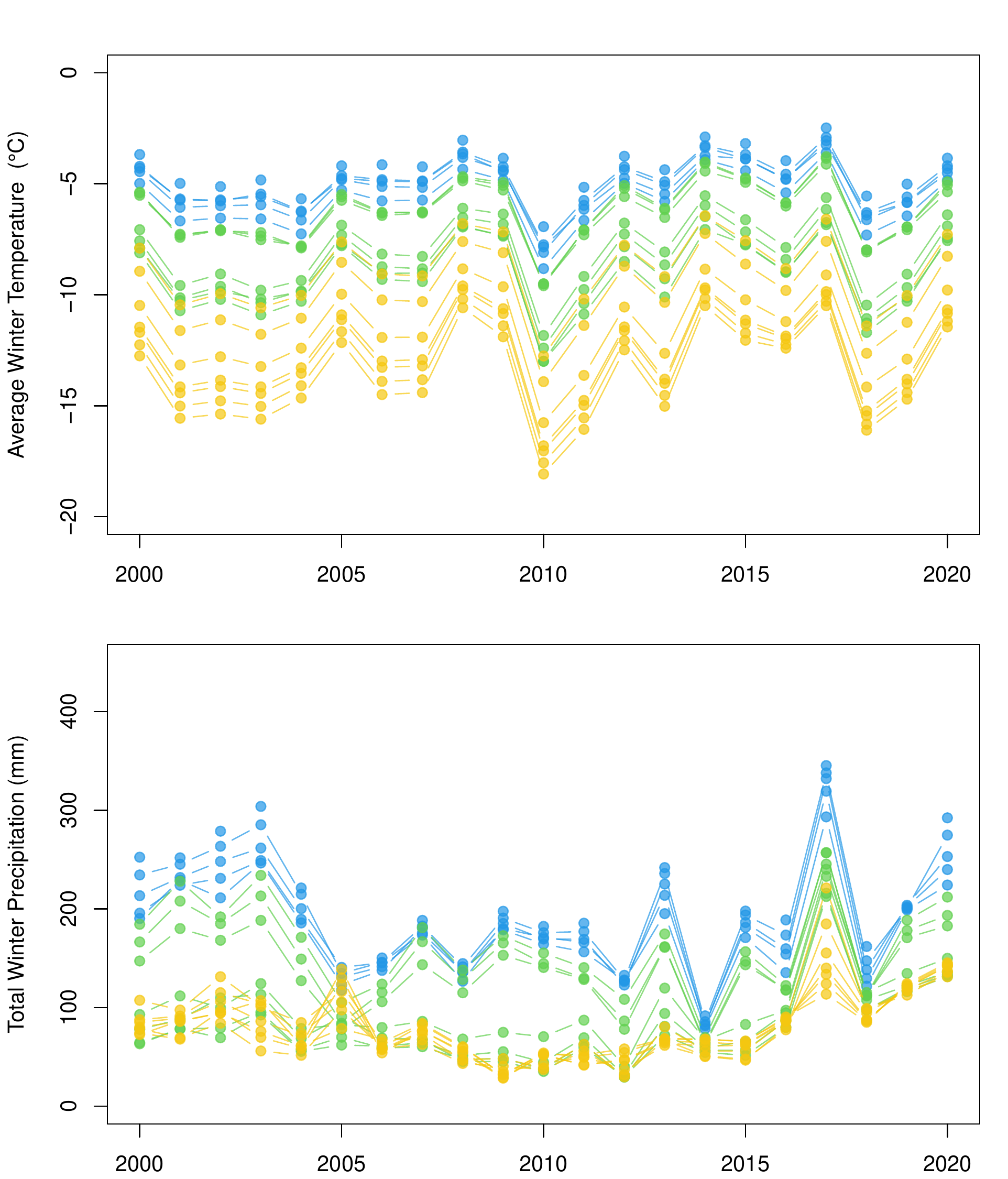}
    \caption{Winter mean temperature and winter total precipitation for the 19 stations, color-coded according to their region.}
    \label{fig:winterpatterns}
\end{figure}

\pagebreak
\section{Model coefficients (section 2.1)}

\subsection*{Model II}
\begin{figure}[!hbt]
    \centering
    \includegraphics[width=.47\linewidth]{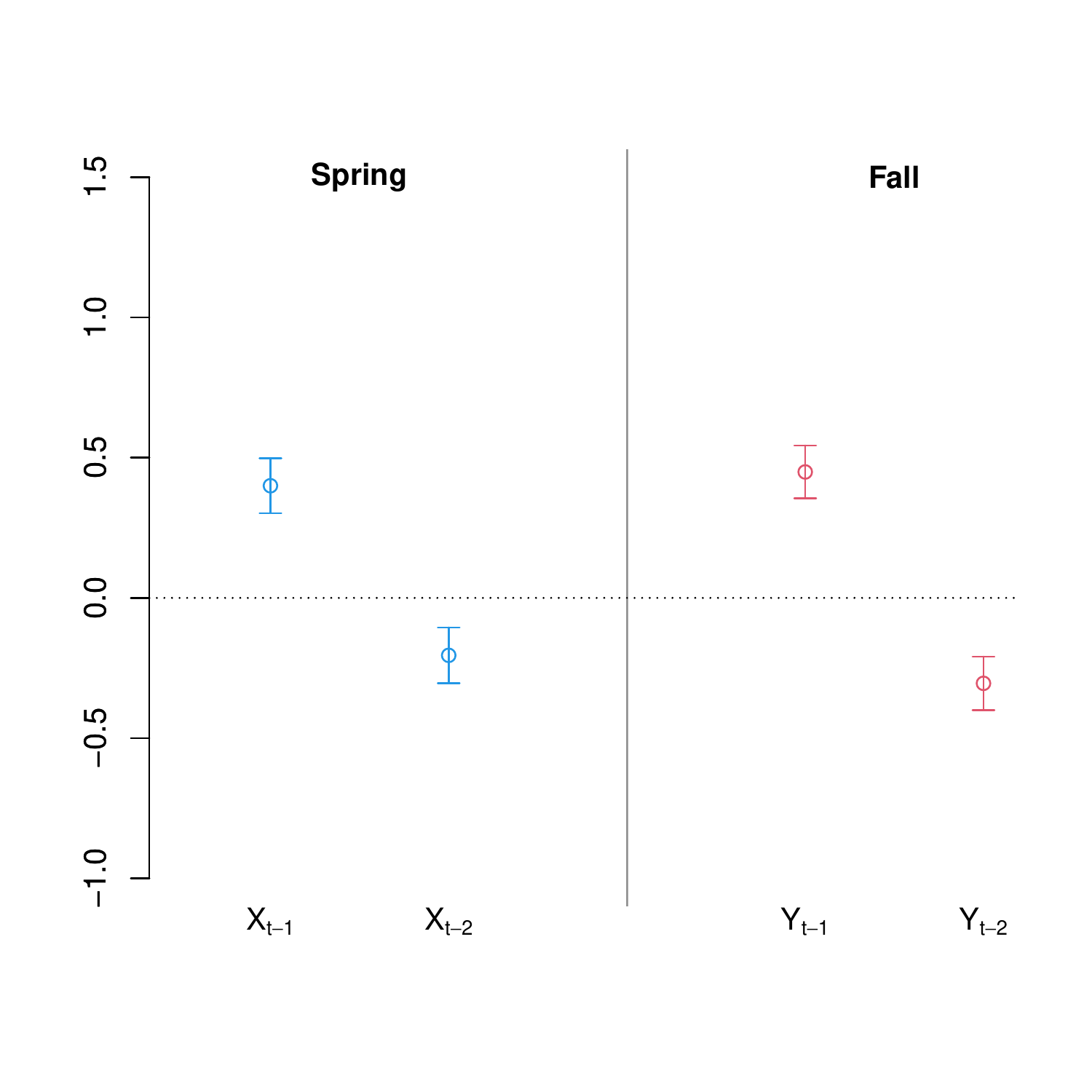}
    \caption{Autoregressive coefficient estimates for model II. Associated $95\%$ credible intervals are represented by the bars. }
  \label{fig:coefsII}
 \end{figure}

\subsection*{Model III}
\begin{figure}[!hbt]
    \captionsetup[subfigure]{labelformat=empty,aboveskip=0pt,justification=centering}
  \centering
  \begin{subfigure}[b]{0.45\linewidth}
    \includegraphics[width=\linewidth] 
    {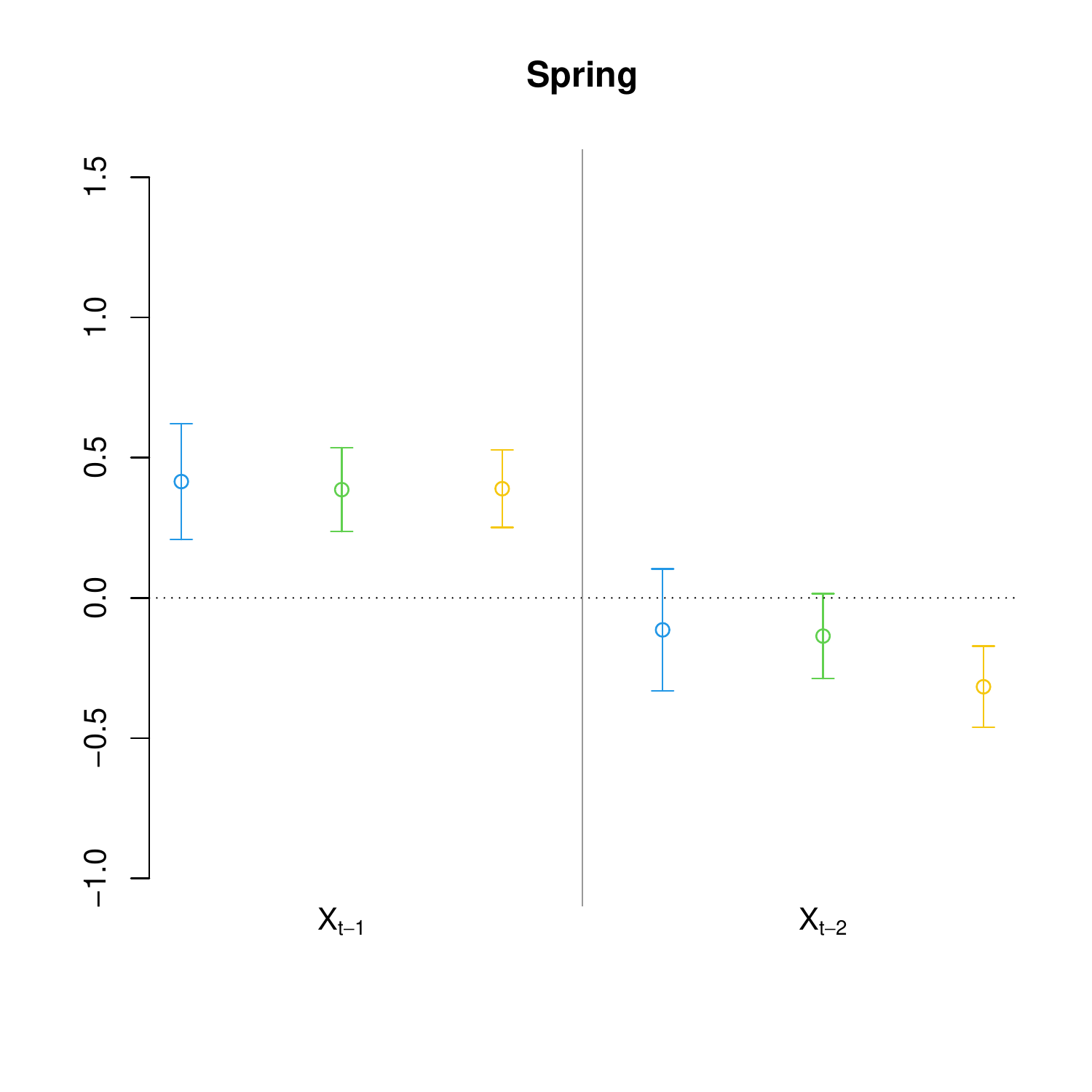}
  \end{subfigure}
  \begin{subfigure}[b]{0.45\linewidth}
    \includegraphics[width=\linewidth] 
    {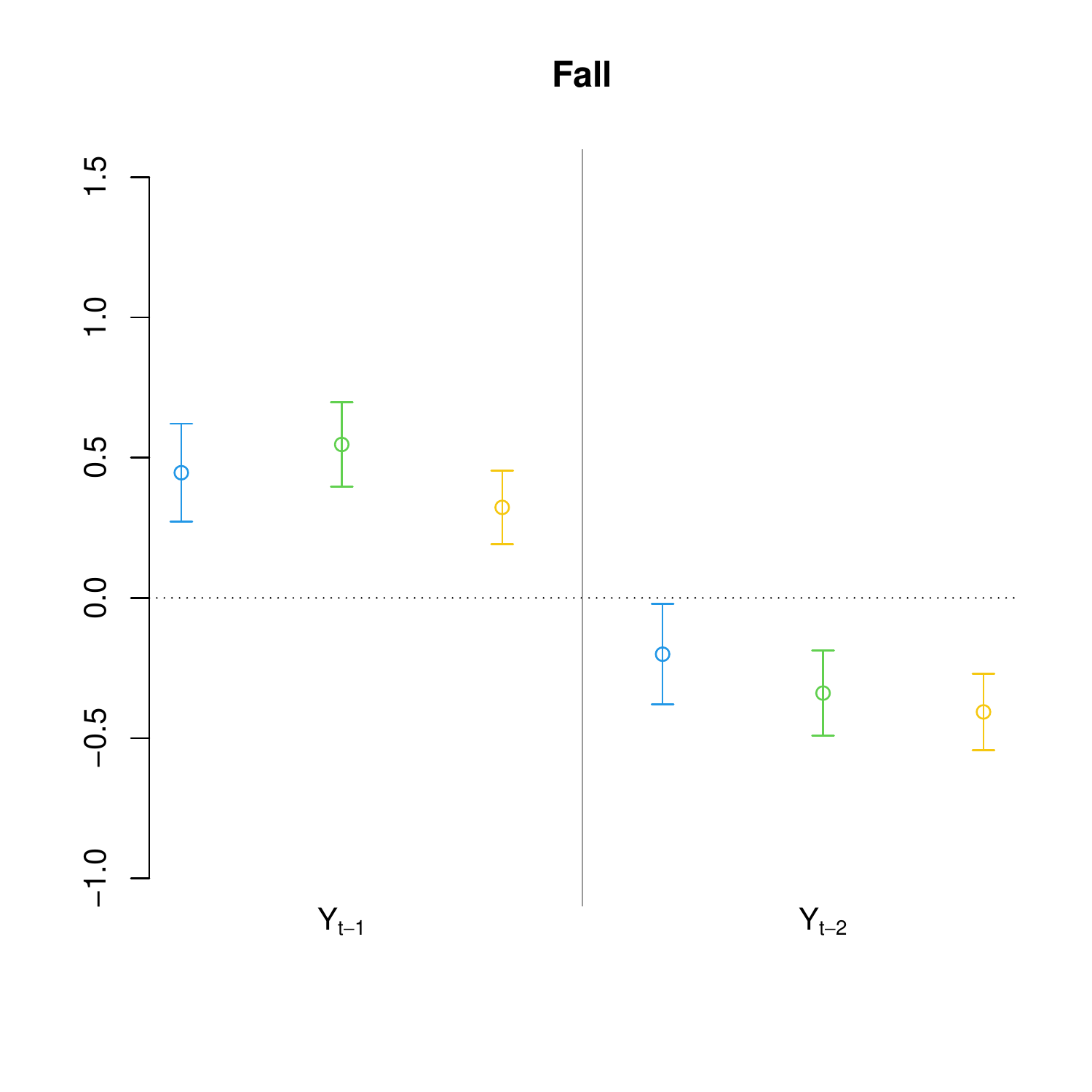}
  \end{subfigure}
    \caption{Autoregressive coefficient estimates for model III. Associated $95\%$ credible intervals correspond to the bars, with the coefficients associated with each region represented with the respective color (blue for $R_1$, green for $R_2$ and yellow for $R_3$). }
  \label{fig:coefsIII}
 \end{figure}

\pagebreak
\subsection*{Model IV} \label{app: modelIVcoefs}

\begin{figure}[!h]
    \captionsetup[subfigure]{labelformat=empty,aboveskip=0pt,justification=centering}
  \centering
  \begin{subfigure}[b]{0.49\linewidth}
    \includegraphics[width=\linewidth] 
    {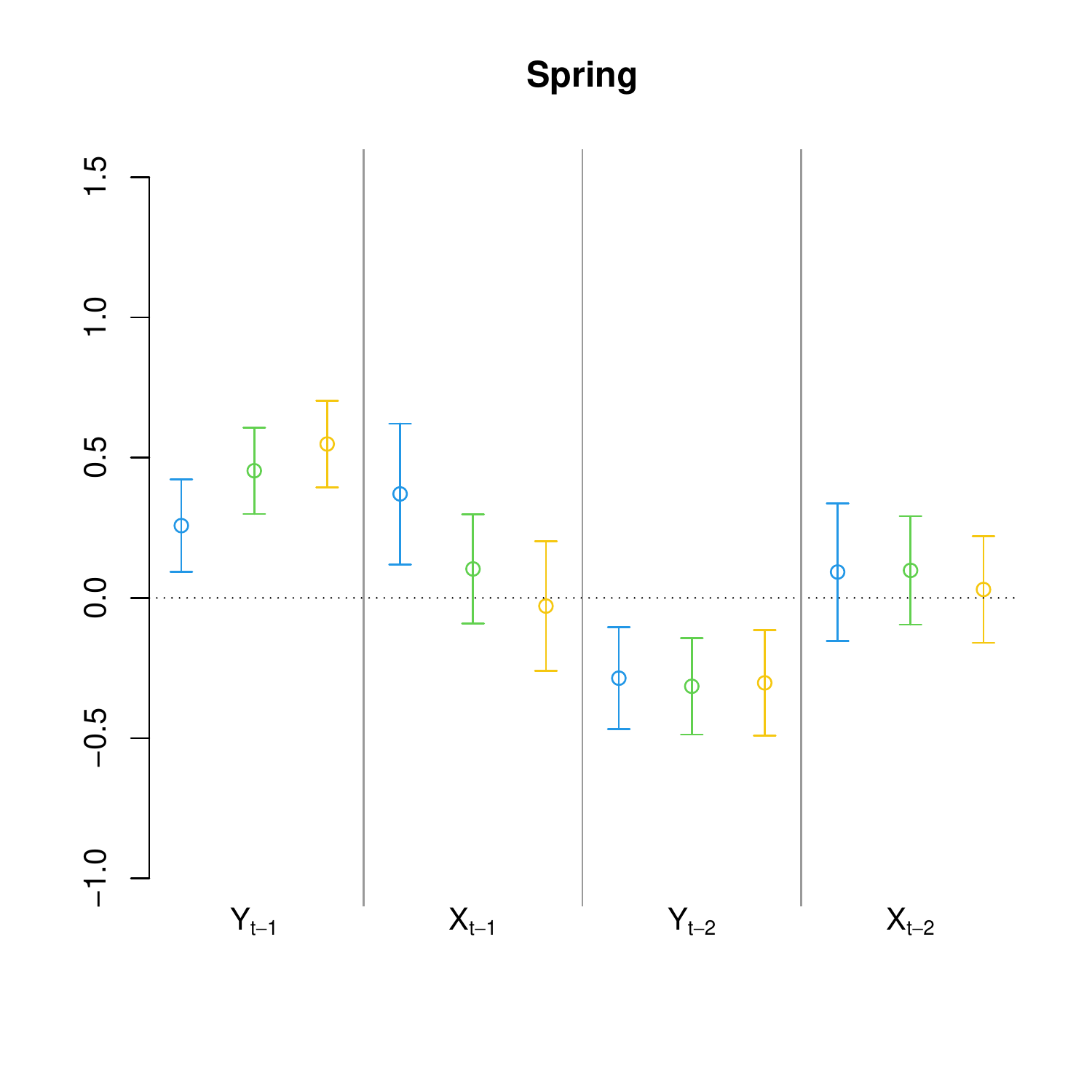}
  \end{subfigure}
  \begin{subfigure}[b]{0.49\linewidth}
    \includegraphics[width=\linewidth] 
    {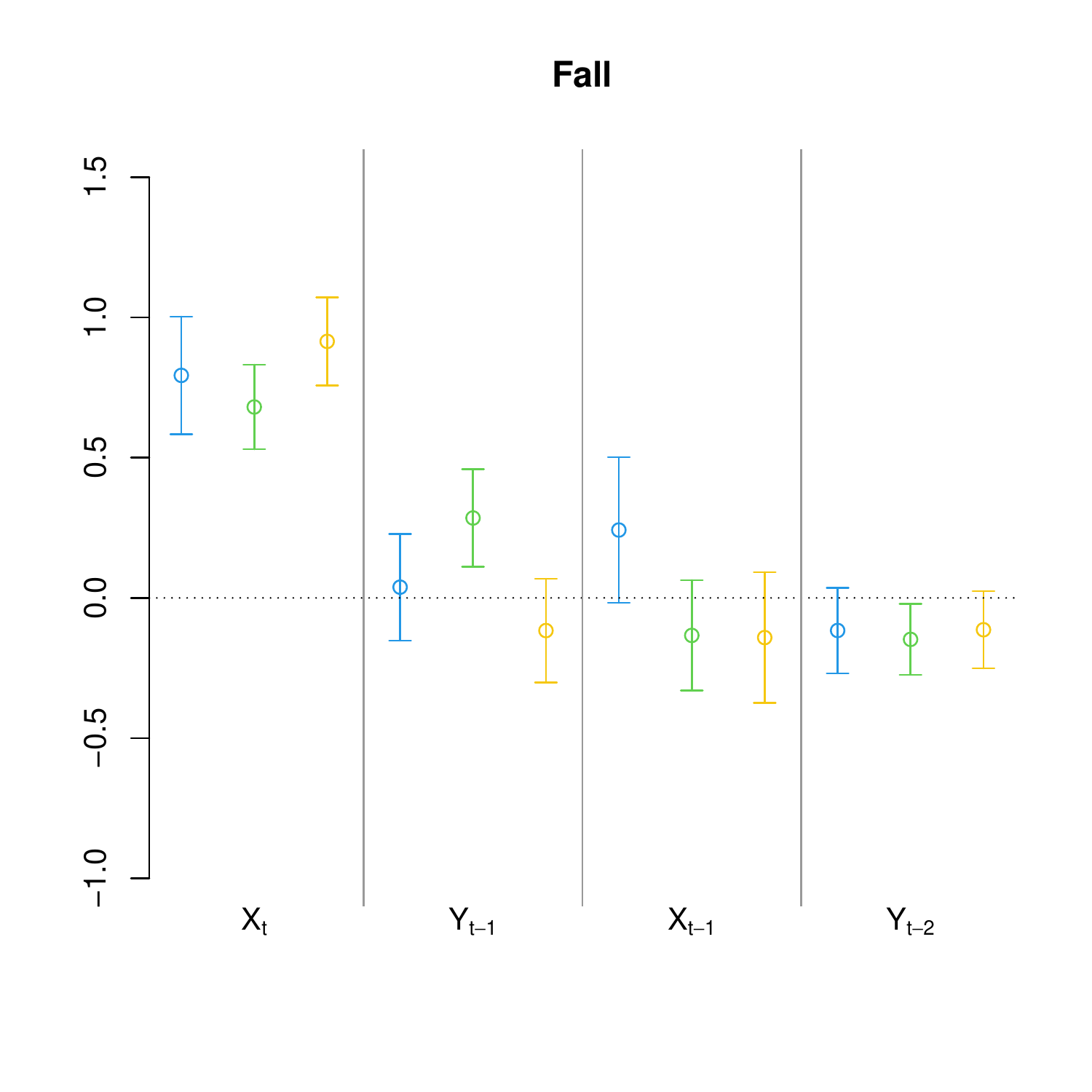}
  \end{subfigure}
    \caption{Autoregressive coefficient estimates for model IV, according to the different seasons. Associated $95\%$ credible intervals correspond to the bars, with the coefficients associated with each region represented with the respective color (blue for $R_1$, green for $R_2$ and yellow for $R_3$). }
  \label{fig:coefsIV}
 \end{figure}

\newpage
\section{Simulations} \label{app:Simulations}

To assess when the observed changes in seasonal correlations can occur, we made a short simulation study. We simulated 100.000 pairs of time series, each pair corresponding to 2 populations with a true process described by model IV (see Methods), with regression parameters approximating those of the real data set (section \ref{app: modelIVcoefs}; parameter values were $\{\beta_1=0.5$, $\beta_2=0.1$, $\beta_3=-0.4$, $\beta_4=0.1$, $\gamma_1=0.8$, $\gamma_2=0.1$, $\gamma_3=-0.1$, $\gamma_4=-0.1\}$. We then compared (a) the true correlation in the noise terms for each pair (which corresponds to a parameter in the simulation; we used r=0.5 for the spring correlation and r=0.3 for the fall correlation), against: (b) the correlations when using year-to-year raw abundances for either spring and fall (model I); and (c) correlations in the residuals of yearly AR(2) for each season separately (model II).

\begin{figure}[!ht]
    \centering
    \includegraphics[width=\textwidth,keepaspectratio]{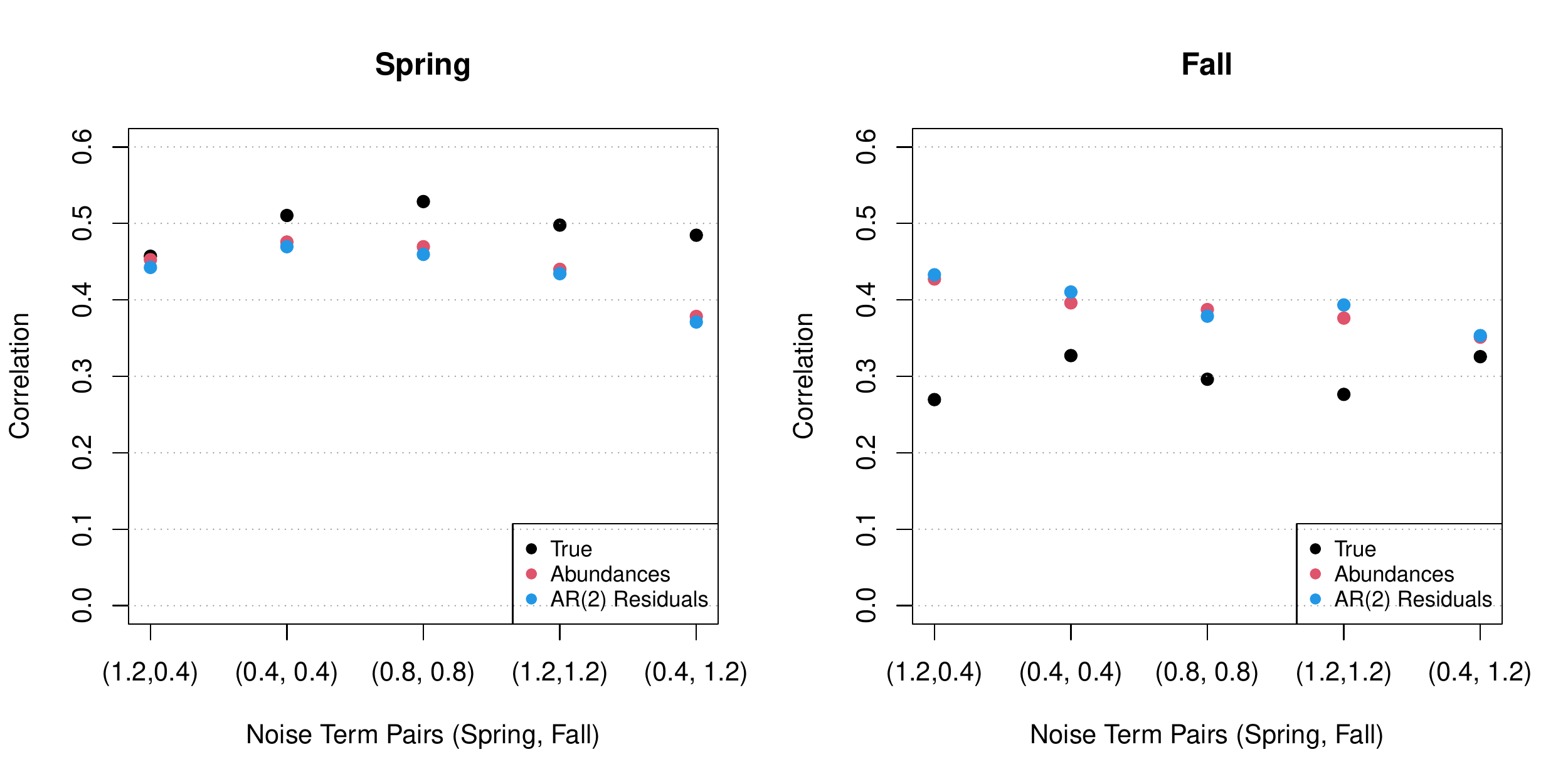}
    \caption{Simulating noise term correlations for either Spring and Fall. X-axis corresponds to the noise terms of either Spring (left) and Fall processes (right). This corresponds to the $\epsilon$ and $\omega$ terms in equations \ref{eq:winter}--\ref{eq:summer}. The mean true correlations in the noise terms (a) are the black dots; the mean correlations in the raw abundances are the red dots; are the mean correlations in the AR(2) model residuals (c) are the blue dots. }
    \label{fig:simulations}
\end{figure}

Figure \ref{fig:simulations} displays the noise correlation estimates for each of the approaches, using different variances of spring and fall noise. We can infer that the decrease we have observed for fall densities (i.e., summer season) happens when the winter noise is as large or larger than the summer noise, which is expected be the case in study system (the estimated variances were 0.86 for spring and 0.86 for fall). If the summer noise is larger than the winter noise, the patterns are reversed compared to the case when winter noise is larger than summer noise . Note that correlations based on annual models can be higher than the true ones. This confirms that the observed changes come from considering the seasonal model (IV) instead of the annual models (I--III), rather than artifacts related to the data. This shows that fluctuating (i.e., seasonal) environments can enhance synchrony, but the exact pattern obtained using annual time series will depend on the noise structure and which season is monitored \citep{Vasseur2007}.

\end{document}